\documentclass{moriond}

\def\be{\begin{equation}}
\def\ee{\end{equation}}
\def\bea{\begin{eqnarray}}
\def\eea{\end{eqnarray}}

\def\Ltopmunu{\ensuremath{\Lambda \to p \mu^{-} \bar{\nu}_{\mu}}\ }
\def\Ltopenu{\ensuremath{\Lambda \to p e^{-} \bar{\nu}_{e}}\ }
\def\Ltoppi{\ensuremath{\Lambda \to p \pi^{-}}\ }
\def\Rmue{\ensuremath{R_{\mu/e}}\ }

\def\BtoDstmunu{\ensuremath{B^0 \to D^{*-} \mu^{+} \nu_{\mu}}\ }
\def\BtoDsttaunu{\ensuremath{B^0 \to D^{*-} \tau^{+} \nu_{\tau}}\ }
\def\BtoDststlnu{\ensuremath{B^0 \to D^{**-} \ell^{+} \nu_{\ell}}\ }
\def\BtoDstDX{\ensuremath{B^0 \to D^{*-} D_{(s)}^{+,0} X}\ }

\def\BtoDststtaunu{\ensuremath{B^-\to D^{**0}\tau^-\bar{\nu}_{\tau}}\ }
\def\BtoDststtaunus{\ensuremath{B^-\to D^{**0}_{1,2}\tau^-\bar{\nu}_{\tau}}\ }
\def\BtoDststmunu{\ensuremath{B^-\to D^{**0}_{1,2}\mu^-\bar{\nu}_{\mu}}\ }
\def\BtoDststtaunuz{\ensuremath{B^-\to D_1(2430)^{0}\tau^-\bar{\nu}_{\tau}}\ }
\def\BtoDststDsst{\ensuremath{B^-\to D^{**0}_{1,2}D_s^{(*)-}}\ }
\def\BtoDststDsXs{\ensuremath{B^-\to D^{**0}_{1,2}D_s^-(X)}\ }
\def\BtoDststDsXz{\ensuremath{B^-\to D_1(2430)^{0}D_s^-(X)}\ }
\def\BtoDstDstX{\ensuremath{B\to D^{*+}D^{*-}(X)}\ }
\def\BtoDstDK{\ensuremath{B\to D^{*+}(DK)^-}\ }
\def\BtoDststpipipi{\ensuremath{B\to D^{**0}\pi^-\pi^-\pi^+}\ }
\def\BtoDststfake{\ensuremath{B\to {\rm fake}\ D^{**0}\pi^-\pi^-\pi^+}\ }

\begin{document}
\vspace*{4cm}
\title{Flavour changing charged current decays at LHCb}

\author{Davide Fazzini, on behalf of the LHCb collaboration}

\address{Department of Physics, University of Milano-Bicocca \& INFN \\
Piazza della Scienza, 3, 20126, Milano, Italy}

\maketitle\abstracts{
The Standard Model (SM) predicts the universality of lepton couplings with the electroweak gauge bosons. 
Semileptonic decays of $b$-hadrons provide a powerful framework for testing the SM and probing possible New Physics effects.
In particular, the processes mediated by charged-current interactions benefit from a relatively large branching fractions and theoretically well-controlled hadronic matrix elements.
This contribution presents three recent results from the LHCb experiment: the first measurement of the ratio of branching fractions $\mathcal{R}(D^{**})$ using $B^{-} \to D^{**0} \tau^{-} \bar{\nu}_{\tau}$ decays, the determination of the branching fraction for \Ltopmunu and the extraction of form-factor parameters from \BtoDstmunu decays.
}

\section{Introduction}

The Standard Model (SM) predicts universal electroweak couplings of the gauge bosons to charged leptons, with any deviations coming solely from the different lepton masses. Any violation of this principle, known as Lepton Flavour Universality (LFU), would provide a clear indication of New Physics (NP) beyond the Standard Model.
Semileptonic $b$-hadron decays proceed via charged-current interactions and are dominated by tree level diagrams in the SM, making them theoretically clean. In addition, their relatively large branching fractions make them an ideal framework for testing the SM consistency and searching for NP effects.

\begin{figure}[!htbp]
    \centering
    \includegraphics[width=0.32\linewidth]{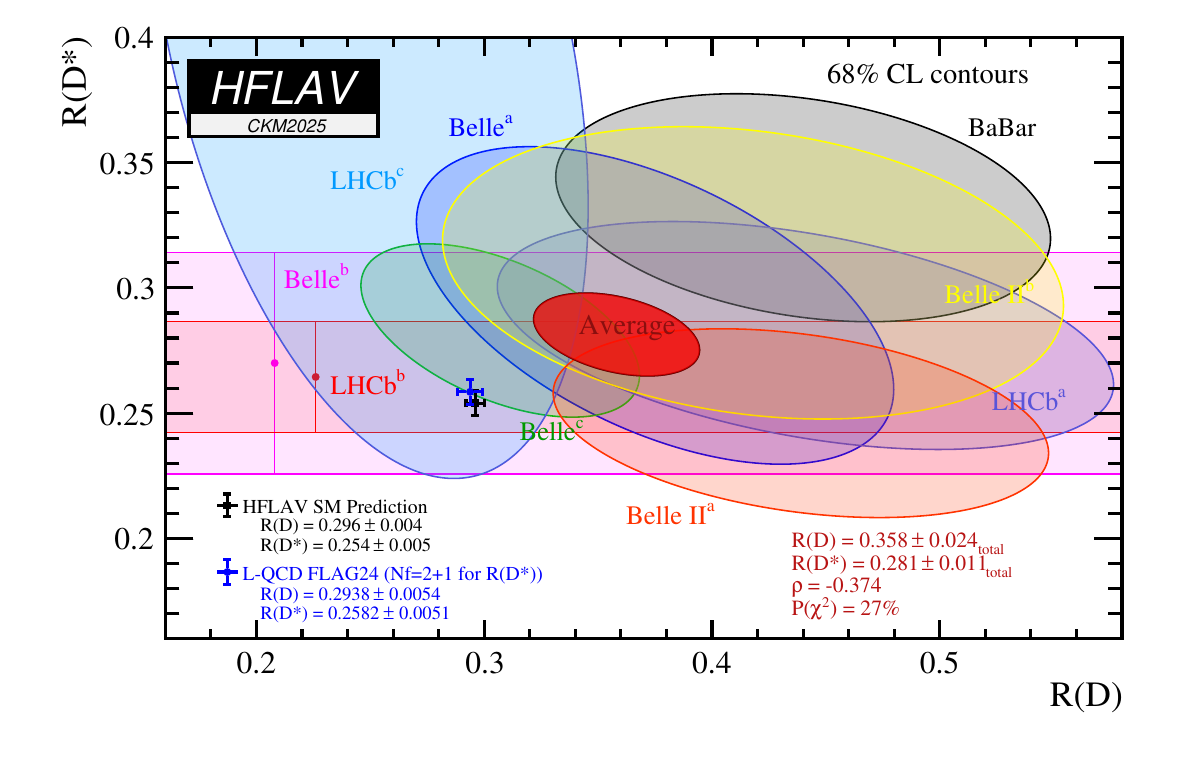}
    \includegraphics[width=0.32\linewidth]{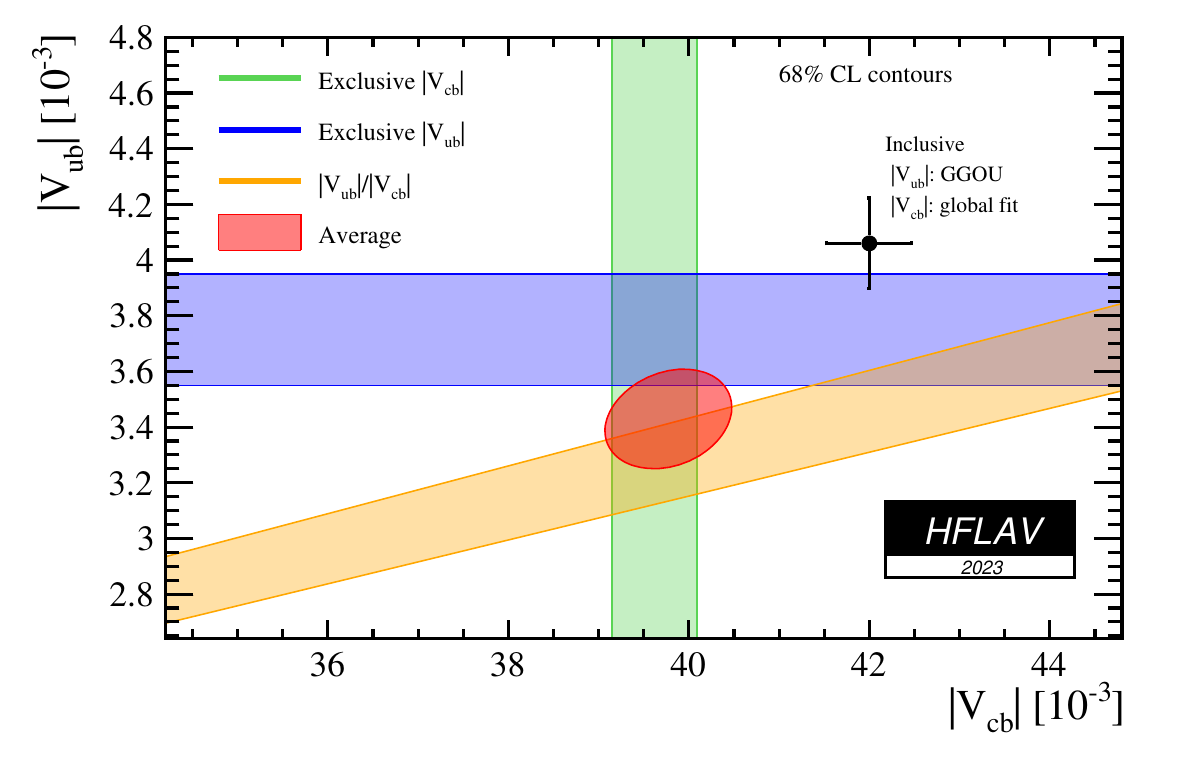}
    \caption{Summary of the measurements of (left) $\mathcal{R}(D)$ and $\mathcal{R}(D^*)$ and (right) $V_{ub}$ and $V_{cb}$ collected by the Heavy Flavor Averaging Group.}
    \label{fig1}
\end{figure}

Interesting discrepancies with respect to the SM predictions have been observed by the  Belle, BaBar and LHCb collaborations in measurement of the ratios $\mathcal{R}(D)$ and $\mathcal{R}(D^*)$, as well as in the determination of the Cabibbo-Kobayashi-Maskawa (CKM) matrix elements $|V_{ub}|$ and $|V_{cb}|$.
In the former case, the discrepancies between the decays with $\tau$ or $\mu$ lepton in the final state are determined through the observable $\mathcal{R}(\mathcal{H}_c)=\frac{\mathcal{B}(\mathcal{H}_b \to \mathcal{H}_c \tau \nu_{\tau})}{\mathcal{B}(\mathcal{H}_b \to \mathcal{H}_c \mu \nu_{\mu})}$.
The current world averages of $\mathcal{R}(D)$ and $\mathcal{R}(D^*)$, depicted in Fig.~\ref{fig1} (left), exhibit a tension of about 3.8$\sigma$ with respect to the SM prediction~\cite{HFLAV}.
In the latter case, a discrepancy at the level of 2-3$\sigma$ is observed between the measurement performed using exclusive and inclusive $b$-hadron decays, shown in Fig.~\ref{fig1} (right).

The LHCb experiment provided significant contributions to this field with various measurements based on semileptonic $b$-hadron decays. Determination of the ratio of branching fractions $\mathcal{R}(D)$ and $\mathcal{R}(D^*)$ using $B^0\to D^{*-}\tau^+\nu_{\tau}$ with leptonic $\tau$ decays~\cite{PhysRevLett.131.111802,PhysRevLett.134.061801}, the measurement of $\mathcal{R}(D^*)$ and the fraction of the longitudinal $D^{*-}$ polarization using \BtoDsttaunu with hadronic $\tau$ decays~\cite{PhysRevD.108.012018} and $\mathcal{R}(\Lambda_c)$ with $\Lambda_b^0\to \Lambda_c^+ \tau^- \bar{\nu}_{\tau}$ decays~\cite{PhysRevLett.128.191803}  have been published recently.

\section{Measurement of the ratio of branching fraction $\mathcal{R}(D^{**}_{1,2})$}

\begin{figure}[!htbp]
    \centering
    \includegraphics[width=0.32\linewidth]{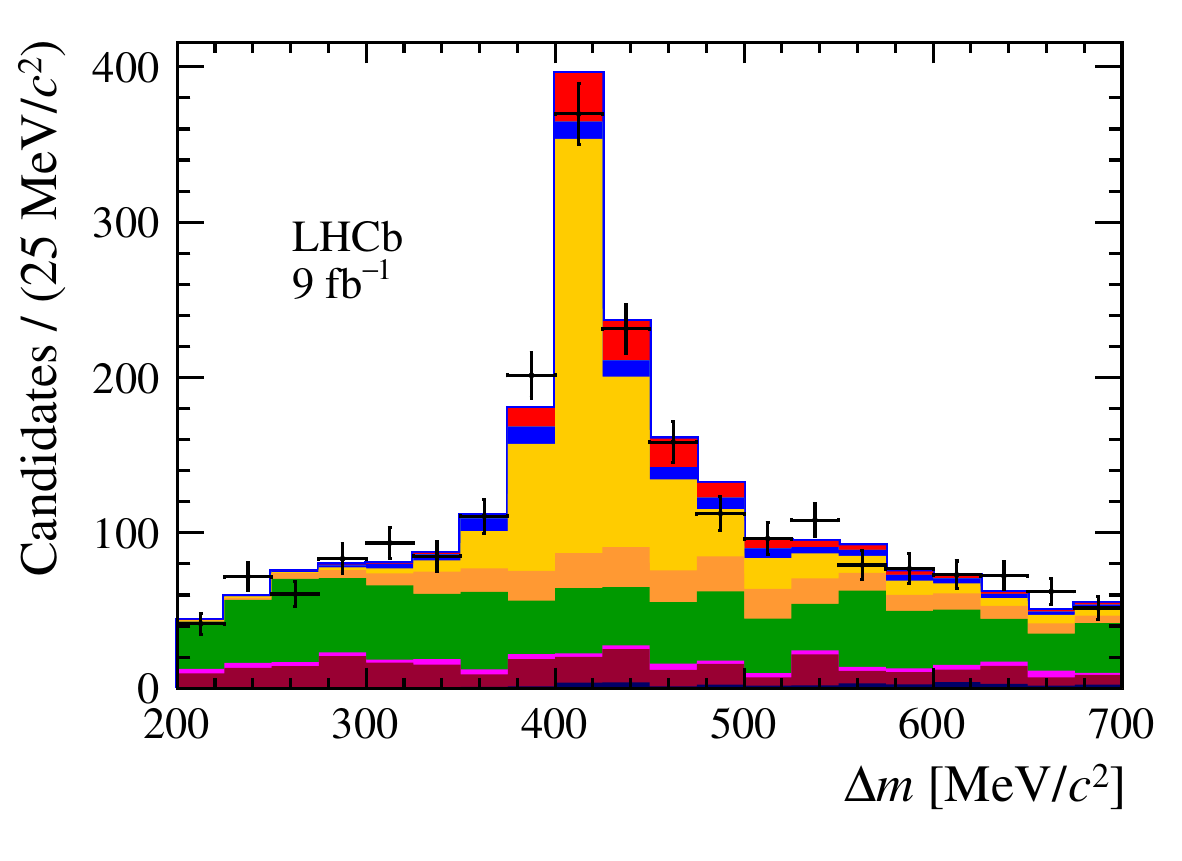}
    \includegraphics[width=0.32\linewidth]{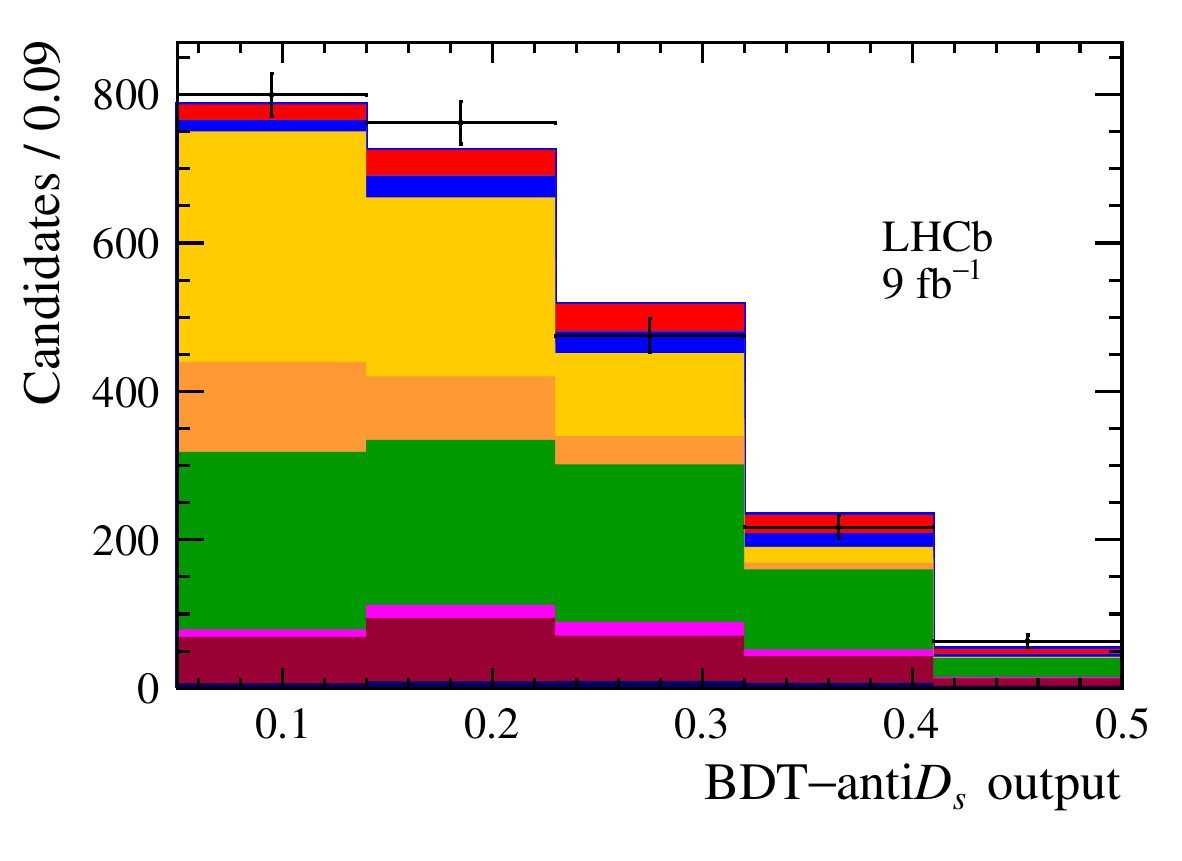}
    \includegraphics[width=0.32\linewidth]{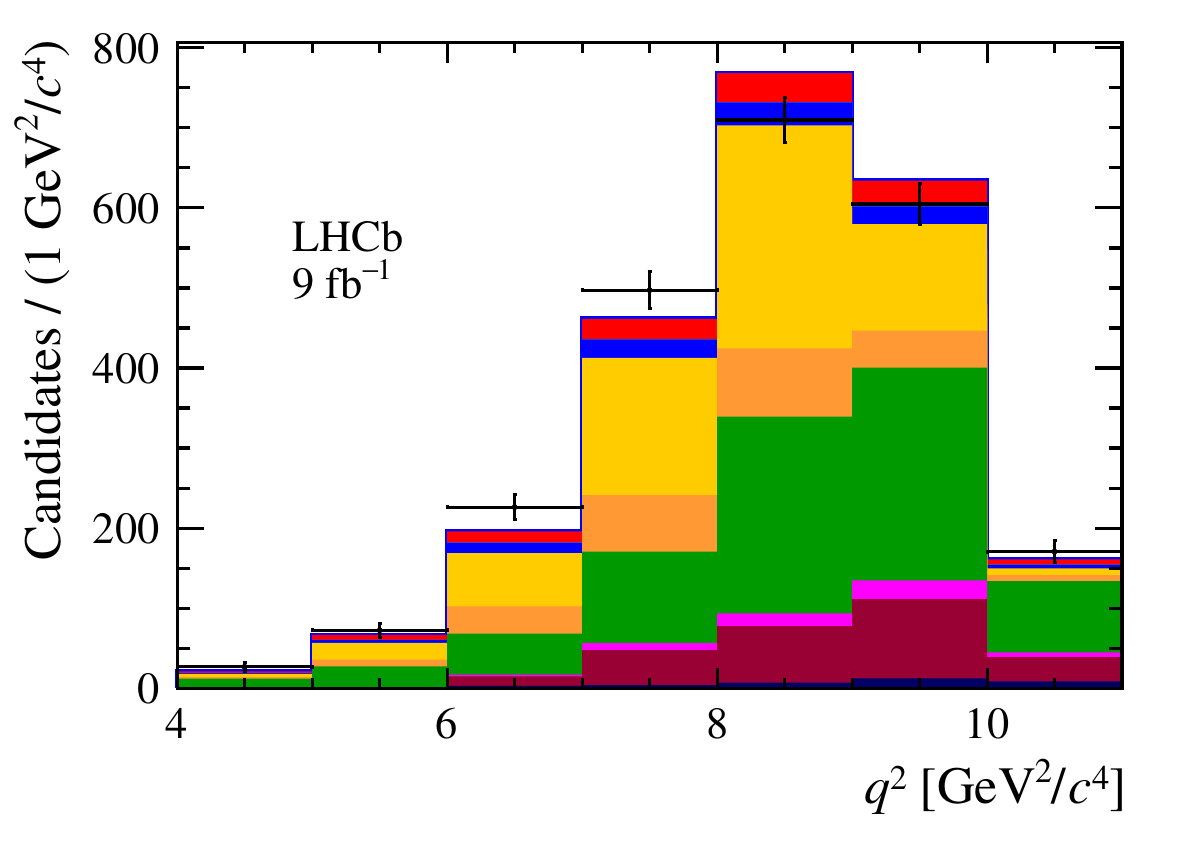}
    \caption{Data distributions (black dots) of the three fit variables with the fit results overlaid. Components: \BtoDststtaunus (red), \BtoDststtaunuz (blue), \BtoDststDsXs (yellow), \BtoDststDsXz (orange), \BtoDstDstX (violet) and \BtoDstDK (bordeaux), \BtoDststfake (green) and \BtoDststpipipi (dark blue).}
    \label{fig2}
\end{figure}

Decays originated from $b$-hadrons into higher excited $c$-meson resonances are backgrounds in measurements of $\mathcal{R}(D^{(*)})$. In the $\mathcal{R}(D^{*})$ measurement using hadronic $\tau$ decays~\cite{PhysRevD.108.012018} the final dataset contained about 3.5\% of \BtoDststtaunu decays.
There are three known $D^{**0}$ resonances that decay into $D^{*+}\pi^{-}$, with masses lying in a range that makes them possible backgrounds for $\mathcal{R}(D^{(*)})$ measurements.
Two of them, $D_1(2420)^0$ and $D_2(2460)^0$, are narrow states with widths of 31 and 47 MeV/c$^2$ while the third one, $D_1(2430)^0$, is a wider state with width of about 315 MeV/c$^2$. In this study, the sum of the two narrow states is referred to as $D^{**0}_{1,2}$.
This analysis exploits the data collected by LHCb in the 2011-2018 period corresponding to an integrated luminosity of 9~fb$^{-1}$, using \BtoDststtaunu candidates reconstructed  with the $\tau$ lepton decaying hadronically.
The \BtoDststDsst mode is chosen as normalization channel, whose branching ratio was measured in an amplitude analysis by the LHCb collaboration~\cite{2024}.
The signal yield is determined through a three-dimensional binned template fit considering the mass difference between the $D^{**0}$ and the $D^{*+}$ candidate ($\Delta m$), the invariant mass of the leptonic system ($q^2$) and the output of the classifier trained to reject the background modes containing a $D_s^+$ meson (BDT-anti$D_s$).
The results of the template fit are shown in Fig.~\ref{fig2} and the signal yield for the $D^{**0}_{1,2}$ mode is determined to be $123\pm23$, representing the first evidence for this decay mode with a significance of 3.5$\sigma$. Combining the resulting visible branching fraction $\mathcal{B}(\BtoDststtaunus)\times\mathcal{B}(D_{1,2}^{**0}\to D^{*+}\pi^-) = (5.1 \pm 1.7)\times 10^{-4}$ with the value of
$\mathcal{B}(\BtoDststmunu)$ measured by BaBar~\cite{PhysRevLett.103.051803} and Belle~\cite{PhysRevD.107.092003} collaborations, the value of the LFU observable is $\mathcal{R}(D^{**}_{1,2}) = 
0.13\pm0.04$~\cite{rj9h-n4w1},
in agreement with the SM prediction of $0.09\pm 0.02$~\cite{PhysRevD.97.075011}.

\section{Measurement of the branching fraction of \Ltopmunu}

The LFU can be tested also using semileptonic hyperon decays through the $s \to u$ transition, since these modes are sensitive to scalar and tensor NP contributions. The LFU observable $\Rmue=\frac{\mathcal{B}(\Lambda\to p\mu^-\bar{\nu}_{\mu})}{\mathcal{B}(\Lambda\to p e^-\bar{\nu}_{e})}$ is predicted to be $0.153\pm0.008$~\cite{PhysRevLett.114.161802}.
In addition, this decay channel allows to measure the $|V_{us}|$ CKM parameter and allowing the test of the unitary constraint for the first row of the CKM matrix, which currently shows a tension at the level of 2$\sigma$~\cite{PhysRevD.110.030001} when combining all the $|V_{us}|$ measurements with the very precise $|V_{ud}|$ value.

This analysis is performed using the LHCb data collected during the 2016-2018 period and corresponding to an integrated luminosity of 5.4~fb$^{-1}$.
The \Ltoppi mode is chosen as normalization channel, whose branching fraction has been measured as $\mathcal{B}(\Ltoppi)=0.641\pm0.005$~\cite{PhysRevD.110.030001}.
The normalization yield is measured from a maximum-likelihood fit to the $m(p\pi)$ invariant mass distribution using candidates surviving a dedicated selection.
The signal yield is obtained using a binned maximum-likelihood fit in bins of $m_{\rm corr}(p\pi)$ and $m(p\pi)$, depicted in Fig.~\ref{fig2} (left), following the procedure described in Ref.~\cite{lhcbcollaboration2025}.
The extracted branching ratio is $\mathcal{B}(\Ltopmunu) = (1.46\pm0.10)\times 10^{-4}$ with a 
total uncertainty of $6.9\%$~\cite{lhcbcollaboration2025}, representing a factor of two improvement in precision with respect to the previous best result obtained by BESIII~\cite{PhysRevLett.127.121802}.
Combining the measured value of $\mathcal{B}(\Ltopmunu)$ with the value $\mathcal{B}(\Ltopenu)=(8.34\pm0.14)\times 10^{-4}$~\cite{PhysRevD.110.030001}, the LFU ratio \Rmue is found to be $0.175\pm0.012$, compatible with the SM prediction~\cite{bvs9-wcsj}.

\begin{figure}[!htbp]
    \centering
    \includegraphics[width=0.32\linewidth]{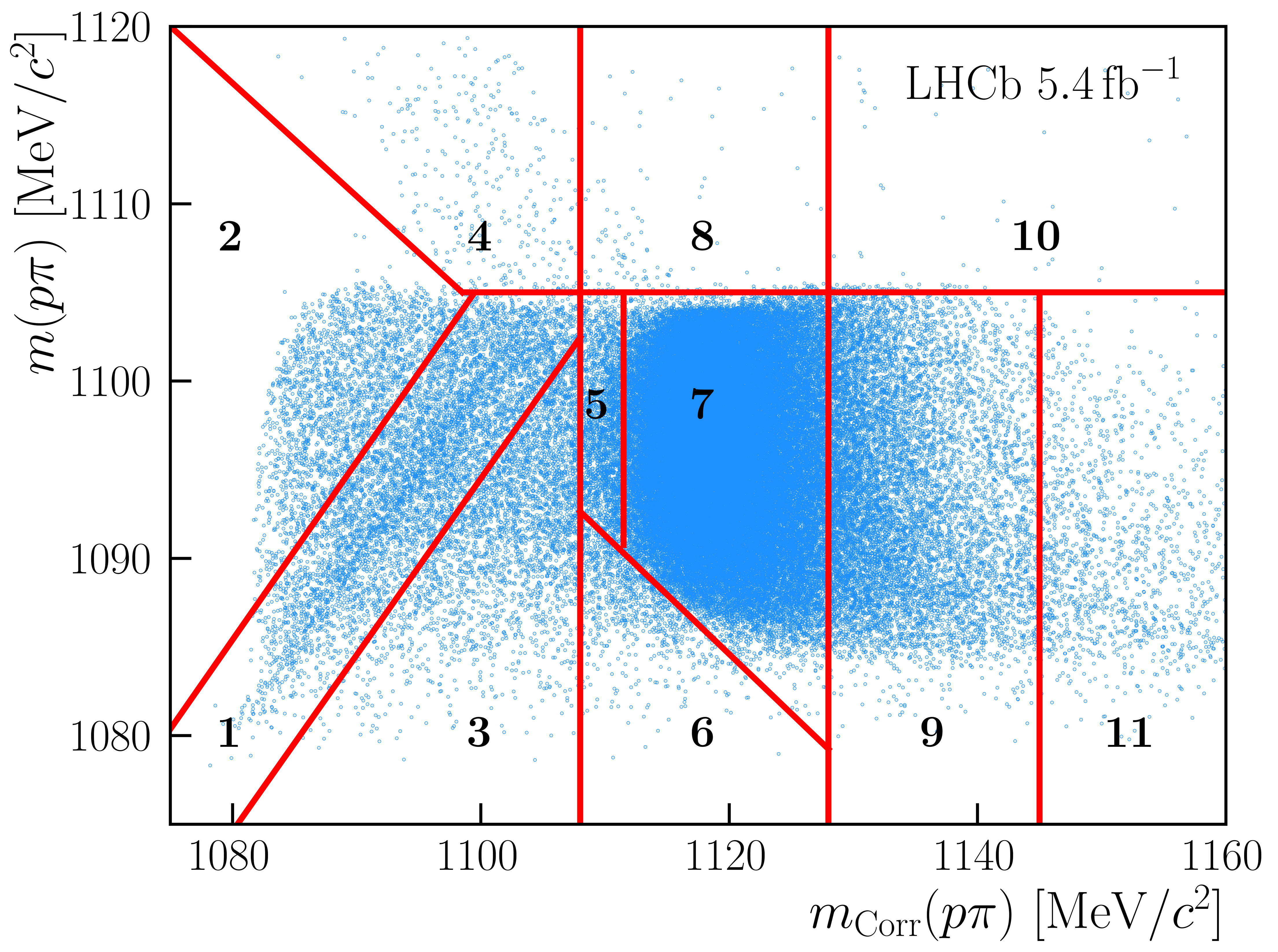}
    \includegraphics[width=0.32\linewidth]{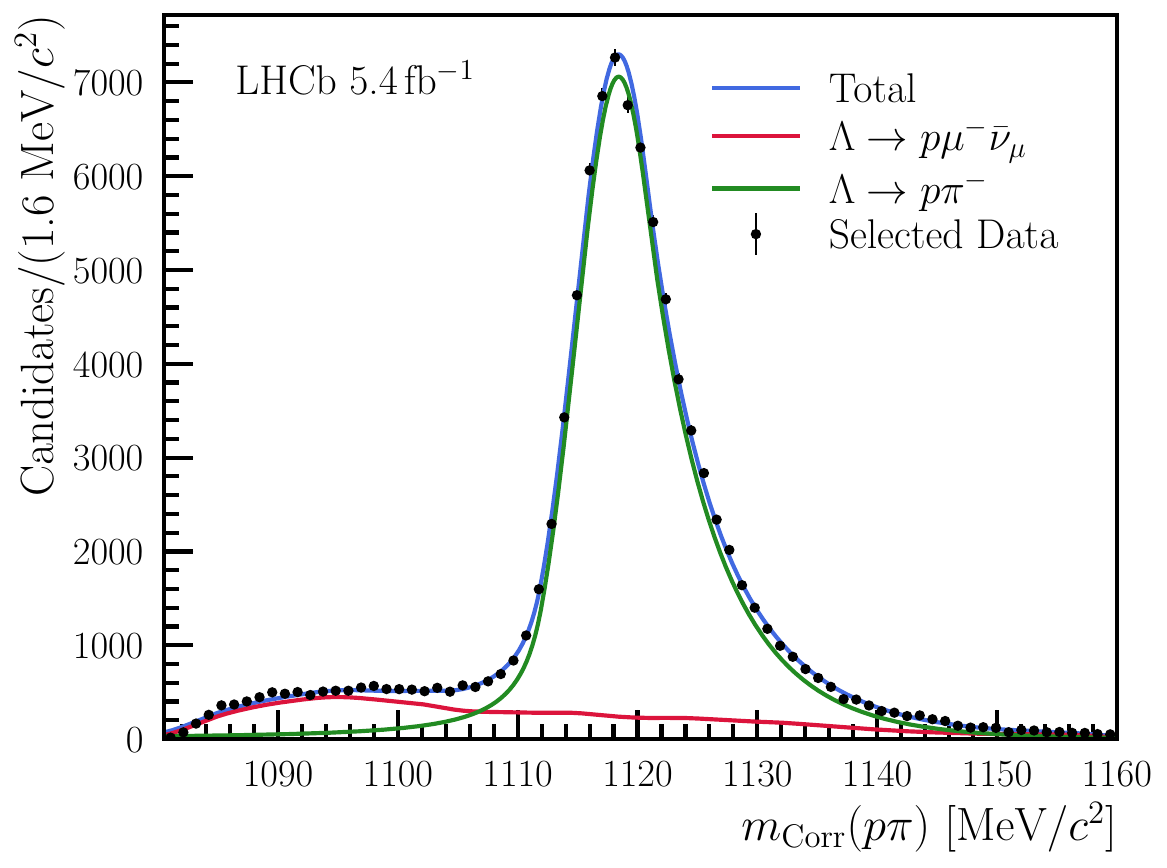}
    \caption{The plots of the (left) the binning scheme in the 2D plane used to perform the binned signal fit and (right) distribution of $m_{\rm corr}$ for candidates surviving the signal selection with the result of the fit overlaid are shown.}
    \label{fig3}
\end{figure}

\section{Measurement of form-factor parameters with $B^{0} \to D^{*-} \mu^{+} \nu_{\mu}$}

\begin{figure}[!htbp]
    \centering
    \includegraphics[width=0.32\linewidth]{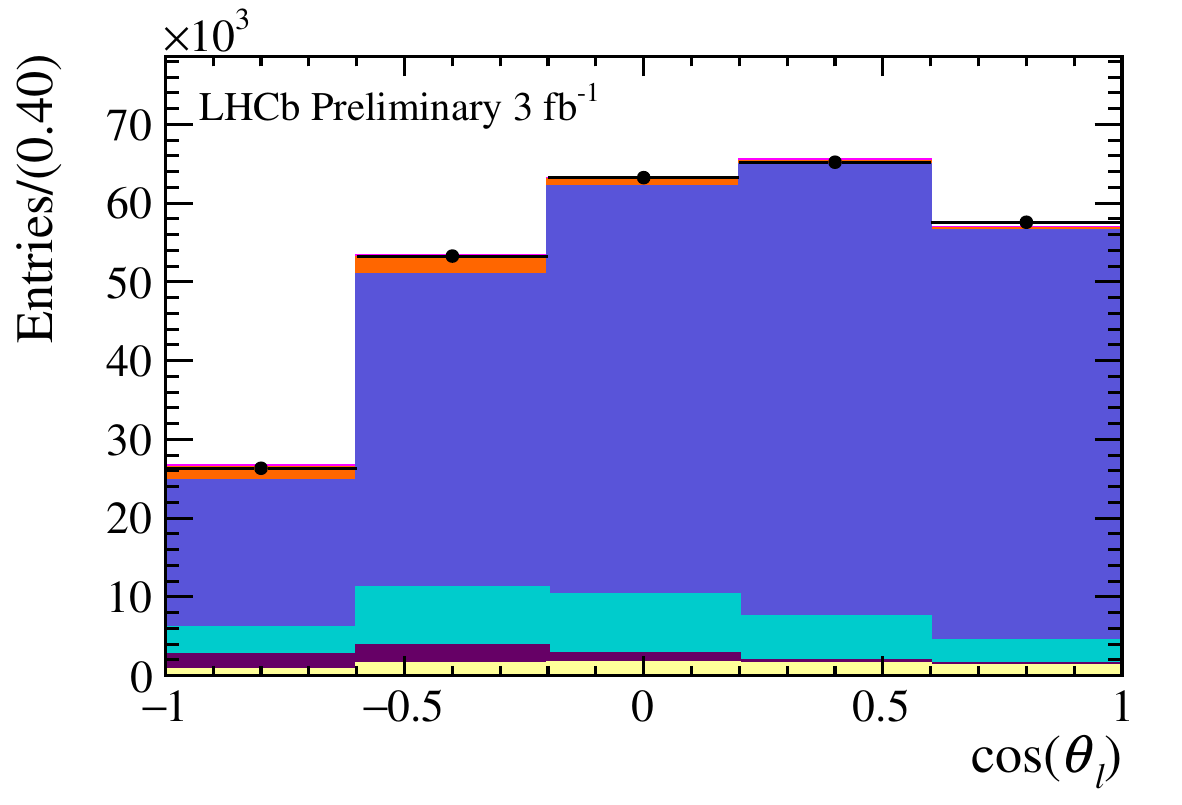}
    \includegraphics[width=0.32\linewidth]{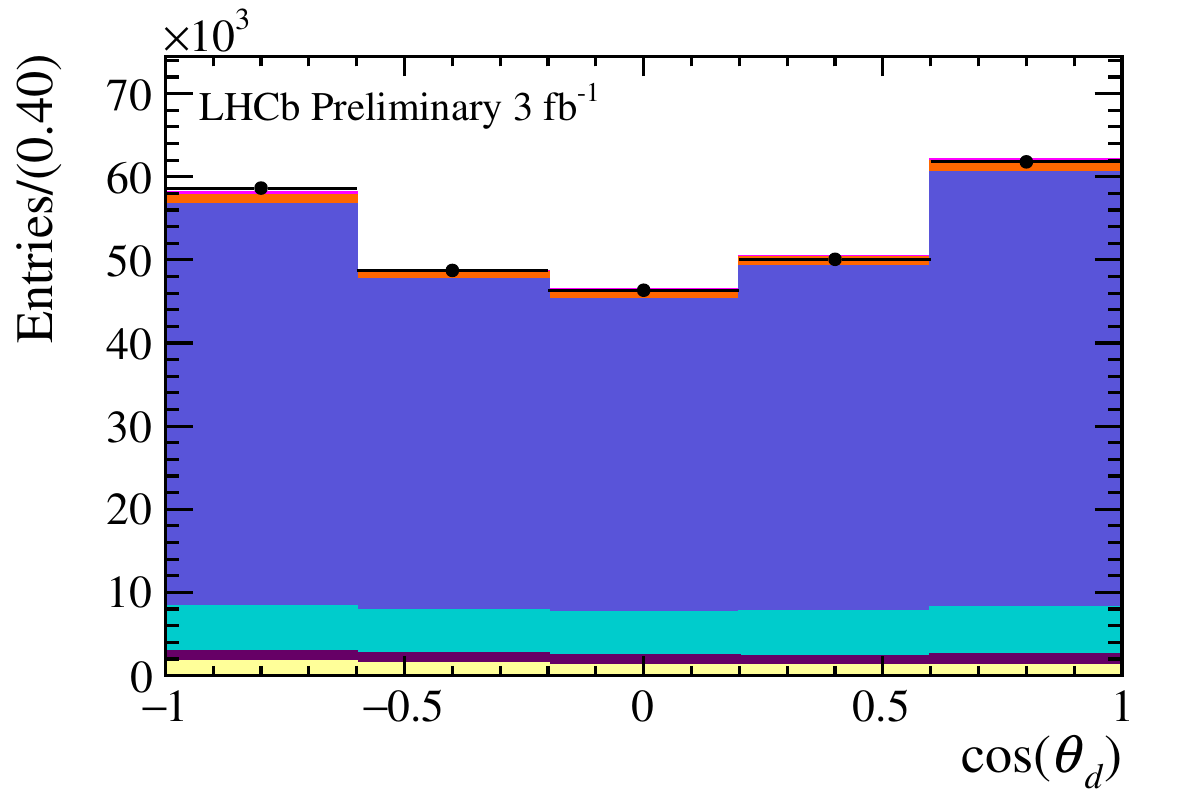}
    \includegraphics[width=0.32\linewidth]{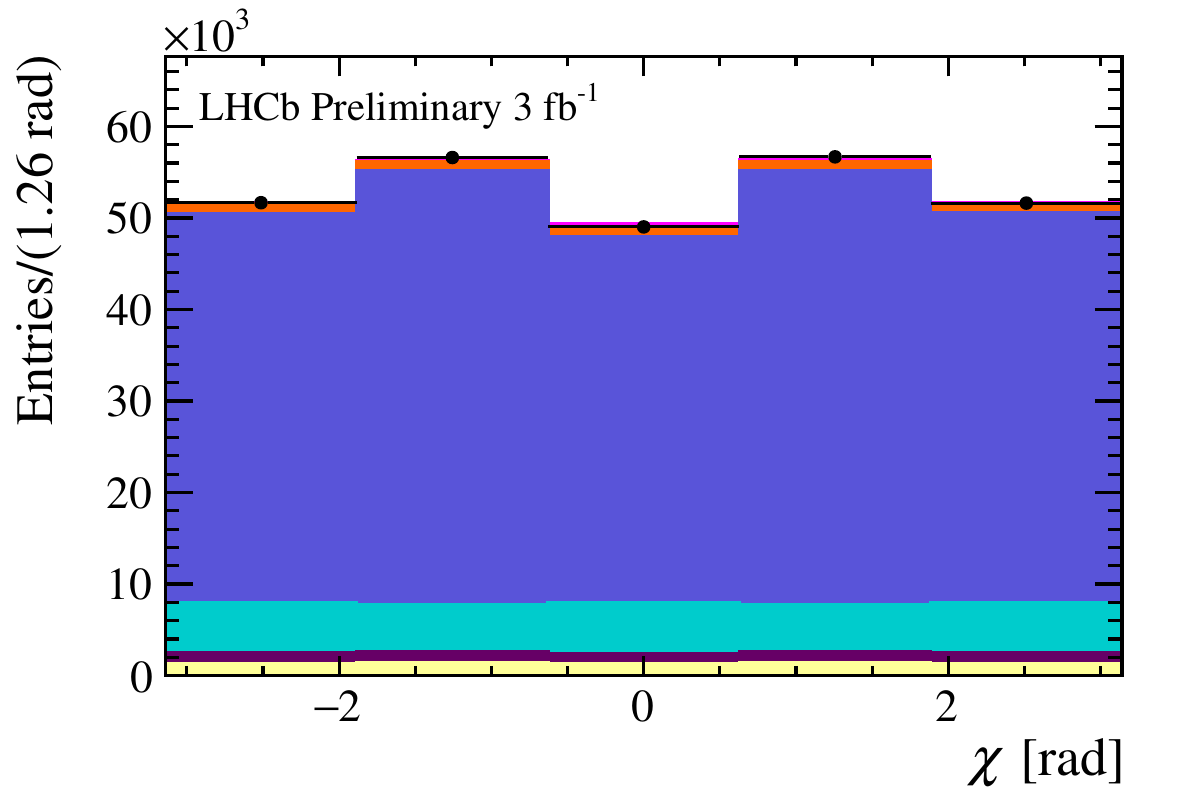}
    \caption{Data distributions of the decay angles (black dots) with the fit results assuming BGL scheme overlaid. Components: \BtoDstmunu (blue), \BtoDsttaunu (bordeaux), \BtoDststlnu (cyan), \BtoDstDX (orange), misID (pink) and combinatorial (yellow).}
    \label{fig4}
\end{figure}

The differential decay rate of the \BtoDstmunu mode can be expressed as function of the decay angles ($\theta_D$, $\theta_{\ell}$ and $\chi$) and the squared momentum transferred to the lepton system ($q^2$), being sensitive to NP effects.
The observable $\theta_D$ ($\theta_{\ell}$) is the angle between the direction of the $D^{0}$ ($\mu$) and the direction opposite the $B^0$ meson, in the $D^{*-}$ ($W^+$) rest frame, while $\chi$ is the angle between the planes formed by the $D^{*-}$ and the $W^+$ decay in the $B^0$ rest frame.
In this analysis the measure of the hadronic form-factors is performed through a five-dimensional binned fit to the three decay angles, $q^2$ and the squared invariant mass missing from the visible system ($m^2_{\rm miss}$).
The analysis is performed using the data collected in 2011-2012 and corresponding to an integrated luminosity of 3.0~fb$^{-1}$.
Due to the presence of a neutrino in the final state, the $B^0$ meson momentum is determined up to a quadratic ambiguity, solved choosing a random solution.
The fit to data is performed assuming three different form-factor parameterizations (BGL~\cite{PhysRevD.56.6895}, CLN~\cite{Caprini_1998} and BLPR~\cite{PhysRevD.95.115008}) for the signal mode, whose hadronic parameters are determined. The fit projections using the BGL scheme are depicted in Fig.~\ref{fig4}~\cite{AngularLHCb}.
The Bayesian information criterion (BIC) method~\cite{c4048c8f} is used to optimize the order of the BGL coefficient series truncation.
The measured values of the BGL parameters satisfy the unitarity constraint.
The measured form-factor values are in agreement between the three models and with the most recent lattice QCD calculations, while achieving improved precision.

\section{Conclusions}

These proceedings present the first measurement of the ratio of branching fraction $\mathcal{R}(D^{**})$ using $B^{-} \to D^{**0} \tau^{-} \bar{\nu}_{\tau}$ decays, the world best determination of the branching fraction for \Ltopmunu and the first LHCb angular analysis with \BtoDstmunu decays extracting the form-factor parameters in CLN, BGL and BLPR parameterizations.
These results confirm the central role of the LHCb experiment in the flavour physics program, offering complementary information to the $B$-factories. 

\section*{References}
\bibliography{main}

\end{document}